\documentclass[aps,prl,twocolumn,groupedaddress,showpacs,showkeys]{revtex4-1}

\usepackage{amsfonts} 
\usepackage{graphicx}
\usepackage{bm}

\begin{document}


\title{Dynamical instabilities of spectroscopic transitions in dense resonant media}




\author{R.A. Vlasov}
 \email{r.vlasov@dragon.bas-net.by}
 \author{A.M. Lemeza}%
 \email{azemel@gmail.com}
 \affiliation{B.I. Stepanov Institute of Physics of the National Academy of Science of Belarus, 68 Nezavisimosti Ave., Minsk, 220072 Belarus}
\author{M.G. Gladush}
 \email{mglad@isan.troitsk.ru}
 \affiliation{Institute for Spectroscopy of the Russian Academy of Sciences, 5 Fizicheskaya Str., Troitsk, Moscow region, 142190 Russia}


\date{\today}

\begin{abstract}
We consider the influence of the near dipole-dipole interaction, underlying the local field, on the dynamics of two-level and V-type three-level atoms exposed to cw-laser radiation. The dipole-dipole interaction between the V-type three-level atoms is shown to give rise to the Poincar\'e-Andronov-Hopf bifurcation in steady-state solution of the equations of motion. As a result, the populations of the energy levels periodically vary in time. In the framework of the two-level model, dynamical instabilities are proved to be impossible.
\end{abstract}

\pacs{42.65.Sf, 42.50.Ct, 42.65.Pc}


\maketitle


In optically dense media, where the density of resonant atoms is high enough, resonant atoms can not be considered as independent ones and the theoretical models should be corrected for the atom-atom interaction. This interaction may include a number of processes of different nature and leads to complex response of media to an external field. One of such processes is the near dipole-dipole (NDD) interaction which becomes more conspicuous with increasing the density of resonant atoms.

Following Ref.~\cite{PhysRevA.47.1247}, allowance for the NDD interaction is made by substituting the local field correction (LFC) into the Bloch equations (or the density matrix equation). Formally, LFC consists in the replacement of the macroscopic filed ${\bf{E}}$  in Bloch equation by the local (or effective) one ${{\bf{E}}_{loc}}$,  according to the relation ${{\bf{E}}_{loc}} = {\bf{E}} + 4\pi {\bf{P}}/3$ , where ${\bf{P}}$ is the induced macroscopic polarization. This phenomenological procedure of introducing LFC is equivalent to a consistent derivation of the Bloch equations from the Bogolubov-Born-Green-Kirkwood-Yvon hierarchy for the reduced single particle density matrices of atoms and quantized field modes  \cite{springerlink:10.1134/S1063776111100050}.

The LFC causes the nonlinearity in the Bloch equations, and it is precisely this feature that has enabled one to predict a number of effects and phenomena: the intrinsic optical bistability (IOB) \cite{PhysRevA.34.3917,PhysRevLett.73.1103}; Lorentz shift \cite{PhysRevLett.67.972,Friedberg1973101}; ultrafast intrinsic optical switching \cite{PhysRevLett.68.911}; enhancement of the absorptionless index of refraction and inversionless gain in dense, coherently prepared, three-level atoms \cite{0954-8998-6-4-014,PhysRevLett.70.1421}; enhancement of the spontaneous decay rate \cite{PhysRevLett.85.1851}; modification of the superradiant amplification characteristics \cite{Manassah2001221}; incoherent soliton formation and phase modulation emergence in self-induced transparency \cite{Afanasev:02}; shifts of the Autler-Townes peaks in absorption spectra \cite{PhysRevA.53.1690}; optical switching operation, stable and unstable hysteresis loops, and an auto-oscillation regime of reflectivity \cite{Malyshev:97,Jarque199766}. A review of the effect of NDD interaction on laser dynamics is given in \cite{Cabrera-2004}.

Among the most dramatic examples of the NDD effect, dynamical oscillations appear to be of particular interest in theoretical and practical terms. The present report is just devoted to this problem. More specifically, our principal concern is with dynamical instabilities developing in the dynamics of purely spectroscopic transitions, leaving aside propagation, boundary and retardation effects as well as laser regimes; this offers a fresh approach to the problem we have touched on. To put it differently, the motivation of the research here would be expressible as a question: Where and how (or under what conditions) could the NDD interaction result in dynamical instabilities when dealing only with spectroscopic transitions?

The search for relevant conditions, if any, is supposed to answer the question posed. For this purpose we consider a thin layer of optically dense medium where the resonant atoms are modeled by two-level and three-level (V-type) quantum systems. Such model enables us to neglect the retardation effects, far-zone part of the local field generated by the dipoles as well as collective radiative relaxation processes (see Refs.\cite{PhysRevA.34.3917,PhysRevA.43.3845}). Based on these assumptions, we analytically prove the impossibility of the dynamical instabilities in the framework of the two-level atom model and show numerically that the instability is to be realizable in a certain range of the parameters of three-level V-type atom model and discuss the experimental condition for the effect observation.

It is worth noting that in the case of optically dense V-type media, the first step in studies of the NDD effect seemed to be taken by the authors of Ref.\cite{springerlink:10.1140/epjd/e2003-00080-2}. However we have taken a further look at the properties of the V-system influenced by the NDD interaction to better appreciate the significance of the NDD effects while arriving at instabilities.

The evolution of the density matrix generalized to the NDD interaction of resonant atoms in the rotating wave approximation can be written as
\begin{equation}\label{eq:01}
i{{d\rho } \over {dt}} = \left[ {{H_0} + {V_A} + {V_C},\rho } \right] + i\Gamma \left( \rho  \right),
\end{equation}
where the ${H_0}$ is the Hamiltonian of free atom,  ${V_A}$ is the operator of the interaction of an atom with the electromagnetic field, ${V_C}$ is the collective operator which results from the NDD interaction, $\Gamma \left( \rho  \right)$ is the relaxation operator. A rigorous derivation of Eq.(\ref{eq:01}) is given in Refs.~\cite{springerlink:10.1134/S1063776111100050}.

Let us assume that the number of levels of a single atom is two. Then the operator in equation (\ref{eq:01}) in the matrix representation has the form \cite{springerlink:10.1134/S1063776111100050}:
\begin{eqnarray}\label{eq:02}
{V_A} =  - \left( {\matrix{
   0 & \Omega_{12}   \cr
   {{\Omega _{21}}} & 0  \cr
 } } \right), \quad
{V_C} =  - \omega_{L}^{21} \left( {\matrix{
   0 & {{\rho _{12}}}  \cr
   {{\rho _{21}}} & 0  \cr
 } } \right),  \nonumber \\
{H_0} = \left( {\matrix{
   0 & 0  \cr
   0 & \Delta_{21}   \cr
 } } \right), \quad
\Gamma \left( {{\rho }} \right) = - {\Gamma_{\|}}\left( {\matrix{
   {-{\rho _{22}}} & {  \gamma {\rho _{12}}}  \cr
   {  \gamma {\rho _{21}}} & {  {\rho _{22}}}  \cr
 } } \right),
\end{eqnarray}
where ${{\Delta}_{21}} =  {\omega_{21}} - {\omega _0}$ is the detuning of the resonant transition frequency ${\omega _{21}}$  from the electromagnetic field frequency ${\omega _0}$, $\Gamma_{\|}\equiv \Gamma_{12}  =
4{\left| {\mbox{\boldmath$\mu$}_{21}} \right|^2}{k^{3}_{A}} l_{1}\left(n\right)/\left( {3\hbar } \right)$
is the relaxation constant (spontaneous emission rate), $\hbar$ is the Plank's constant, $k^{3}_{A}={\omega _{21}}/c$, $c$ is the velocity of light in vacuum, ${\Omega _{12}} = \Omega _{21}^* = {{{\mbox{\boldmath$\mu$}_{21}}} }\left(n\right){\bf{E}_0}l_{2}/ {\hbar}$  is the Rabi frequency, ${\bf{E}_0}$ is the slowly varying amplitude of electromagnetic field ${\bf{E}}={\bf{E}_0}e^{-i \omega _0 t} + c.c.$,   ${\mbox{\boldmath$\mu$}_{21}}$ is the transition dipole moment, $\omega_{L}^{21}= {{4\pi } }{n_A}{\left|{\mbox{\boldmath$\mu$}_{21}}\right|}^{2}l_{3}\left(n\right)/\left({3\hbar }\right)$ is the LFC parameter (Lorentz frequency shift), ${n_A}$ is the density of atoms. The correction factors $l_{i}\left(n\right)$ take into account the influence of dielectric media (host) with refractive index $n$ and depends on the type of the impurity \cite{springerlink:10.1134/S1063776111100050}. $\gamma = \Gamma_{\bot}/\Gamma_{\|}$ is the dephasing rate constant in units of the population relaxation rate constant ($\gamma= 1/2$ for the natural linewidth broadening only).

On substitution of operators (\ref{eq:02}) into (\ref{eq:01}) and transformation of the complex equations to real ones the nonlinear system of Bloch equations reads
\begin{eqnarray}\label{eq:03}
\dot{r}_{12}&=&\left( {\beta - \delta  - 2\beta{r_{22}}} \right){r_{21}} + {\Phi _{21}}\left( {1 - 2{r_{22}}} \right) - \gamma {r_{12}}, \nonumber \\
\dot{r}_{21}&=&-\left( {\beta - \delta  - 2\beta{r_{22}}} \right){r_{12}} - {\Phi _{12}}\left( {1 - 2{r_{22}}} \right) - \gamma {r_{21}}, \nonumber \\
\dot{r}_{22}&=&2\left( {{\Phi _{21}}{r_{12}} - {\Phi _{12}}{r_{21}}} \right) - {r_{22}},
\end{eqnarray}
where we denote ${r_{12}} = {\mathop{\rm Re}\nolimits} [{\rho _{12}}]$, ${r_{21}} = {\mathop{\rm Im}\nolimits} [{\rho _{12}}]$, ${r_{ii}} = {\rho _{ii}}$, ${\Phi _{12}} = {\mathop{\rm Re}\nolimits} [{\Omega _{12}}]/\Gamma_{\|}$, ${\Phi _{21}} = {\mathop{\rm Im}\nolimits} [{\Omega _{12}}]/\Gamma_{\|}$, $\delta=\Delta_{21}/\Gamma_{\|}$, $\beta=\omega_{L}^{21}/\Gamma_{\|}$  and exclude one equation due to the normalization condition in the closed two-level system $Tr(\rho) = 1$. The derivatives with respect to the dimensionless time $\tau=\Gamma_{\|}t$ are designated as ${\dot{r}}_{ij}$.

To perform the stability analysis of the steady state solutions ${{\tilde r}_{ij}}$ of (\ref{eq:03}), it is necessary to linearize the system in the vicinity of ${{\tilde r}_{ij}}$, construct the Jacobi matrix $J\left( {{{\tilde r}_{ij}}} \right)$ and examine its eigenvalues. Then the eigenvalue problem is reduced to the solution of the characteristic polynomial with respect to the eigenvalues $\lambda_{i}$:
\begin{equation}\label{eq:05}
\det \left( {J\left( {{{\tilde r}_{i,j}}} \right) - \lambda I} \right) = {a_3}{\lambda ^3} + {a_2}{\lambda ^2} + {a_1}\lambda  + {a_0} = 0,
\end{equation}
where $I$ is the $3\times3$ identity matrix, the coefficients $a_i$ being cumbersome expressions that are not represented here.

In the search for instabilities, there is a need to find conditions when the real parts of eigenvalues ${\lambda _i}$ become positive. So, when the complex eigenvalue intersects the imaginary axis of the complex plane, we can represent the eigenvalues as $\lambda  = i\zeta $, where $\zeta$ is real, and substitute them into (\ref{eq:05}). Upon separating the resulting equation into real and imaginary parts and eliminating $\zeta$, we get the conditions whereby the real part of the eigenvalue changes the sign:
\begin{equation}\label{eq:07}
{a_0} = 0 \qquad \text{or } \qquad { {a_0}{a_3} - {a_1}{a_2} } = 0.
\end{equation}
It follows from equation (\ref{eq:05}) that the first condition of (\ref{eq:07}) corresponds to the sign change of real eigenvalue of the Jacobi matrix and leads to the IOB occurrence. In the explicit form it can be expressed as
\begin{eqnarray}\label{eq:08}
{\gamma ^2} + {(\beta - \delta )^2} - 4\beta {\tilde r_{22}}(2(\beta - \delta ) + \nonumber \\  + {\tilde r_{22}}( - 5\beta + 2\delta  + 4\beta{\tilde r_{22}})) = 0,
\end{eqnarray}
where we formally regard a steady value of the population ${\tilde r_{22}}$ of the upper level as an independent variable in interval $0 \le {\tilde r_{22}} < 0.5$ (the condition of impossibility of population inversion in the framework of the two-level atom model). The relation between ${\tilde r_{22}}$ and dimensionless amplitude of Rabi frequency   $\Phi=\sqrt{\Phi_{12}^2 + \Phi_{21}^2} $ is given by steady state solutions (\ref{eq:03}). Roots of equation (\ref{eq:08}) define the critical points of a hysteresis loop of IOB as well as a condition of its occurrence (it turns out to be ${(\beta  - \delta )^2}(\beta + 8\delta ) > 27{\gamma ^2}\beta $).

The explicit form of the second condition (\ref{eq:07}) can be written as
\begin{eqnarray}\label{eq:09}
(({{\tilde r}_{22}}\gamma  & + (1 - 2{{\tilde r}_{22}}){\gamma ^2})(\gamma (1 + \gamma ) + {( \delta - (1 - 2{{\tilde r}_{22}})\beta)^2}) + \nonumber \\ & +
(1 - 2{{\tilde r}_{22}})({\gamma ^2}(1 + \gamma ) - {{\tilde r}_{22}}\beta \delta ) + \nonumber \\& +
{{\tilde r}_{22}}{\delta ^2} )(1 - 2{{\tilde r}_{22}}) = 0
\end{eqnarray}

Eq.~(\ref{eq:09}) has real roots and predicts the Poincar\'e-Andronov-Hopf bifurcation occurrence (the emergence of a pair of complex eigenvalues with the positive real parts) only outside the interval of physically allowed values of the upper level population ($0 \le {\tilde r_{22}} < 0.5 $) and at $ \Phi^{2}<0$. This implies the absence of dynamical instabilities in the framework of the two-level atom model.

In spite of this, there is a hope to find oscillations (or chaos) in more complex cases of configurations of the atomic energy levels. For our analysis, among all the possible configurations of energy levels of three-level atoms, we choose the V-type level system and perform analysis numerically because of complexity associated with a great number of dynamical variables and parameters.

Let us assume that the three energy levels of a single atom have the V-type configuration with the ground state $\left| 1 \right\rangle $  and two exited states $\left| 2 \right\rangle $  and $\left| 3 \right\rangle $. The transition  $\left| 2 \right\rangle \leftrightarrow \left| 3 \right\rangle $  is forbidden (the transition dipole moment ${\mbox{\boldmath$\mu$}_{23}} = 0$) and the remaining transitions $\left| 1 \right\rangle \leftrightarrow \left| 2 \right\rangle,\left| 3 \right\rangle $ are coupled by one and the same laser field. The transition frequencies obey the condition ${\omega _{32}} <  < {\omega _{21}},{\omega _{31}}$. The absolute values of the dipole transition dipole moments are equal $\mu  = \left| {{\mbox{\boldmath$\mu$}_{21}}} \right| = \left| {{\mbox{\boldmath$\mu$}_{31}}} \right|$  and orthogonal $\left( {{\mbox{\boldmath$\mu$}_{21}} \cdot {\mbox{\boldmath$\mu$}_{31}}} \right) = 0 $ (the quantum interference is absent). It is also anticipated that the stationary electromagnetic field has the frequency ${\omega _0} = {{\left( {{\omega _{21}} + {\omega _{31}}} \right)} \mathord{\left/
 {\vphantom {{\left( {{\omega _{21}} + {\omega _{31}}} \right)} 2}} \right. \kern-\nulldelimiterspace} 2}$. At the chosen parameters the Rabi frequencies, relaxation constants and local field parameters are: $\Omega \equiv {\Omega _{12}} = {\Omega _{13}}$, $ \Delta \equiv {\Delta _{31}} =  - {\Delta _{21}}  $, $  \Gamma_{\|} \equiv {\Gamma _{12}} = {\Gamma _{13}} $, $\omega_{L} \equiv \omega_{L}^{21} = \omega_{L}^{31}$.  In so doing, operators of (\ref{eq:01}) have the form
\begin{eqnarray}\label{eq:10}
H_A^0 &=& \left( {\matrix{
   0 & 0 & 0  \cr
   0 & { - \Delta } & 0  \cr
   0 & 0 & \Delta   \cr
 } } \right), \quad
{V_A} =  - \left( {\matrix{
   0 & \Omega  & \Omega   \cr
   \Omega ^*  & 0 & 0  \cr
   \Omega ^*  & 0 & 0  \cr
 } } \right), \nonumber \\
\Gamma \left( {{\rho}} \right) &=& -\Gamma_{\|} \left( {\matrix{
   {-{\rho _{22}} - {\rho _{33}}} & {  \gamma _{12}{\rho _{12}}} & {  \gamma _{13}{\rho_{13}}}  \cr
   {  \gamma _{12}{\rho _{21}}} & {  {\rho _{22}}} & {  \gamma _{23}{\rho _{23}}}  \cr
   {  \gamma _{13}{\rho _{31}}} & {  \gamma _{23}{\rho _{32}}} & {  {\rho _{33}}}  \cr
 } } \right), \nonumber \\
{V_C} &=&  - \omega _{L}\left( {\matrix{
   0 & {{\rho _{12}}} & {{\rho _{13}}}  \cr
   {{\rho _{21}}} & 0 & 0  \cr
   {{\rho _{31}}} & 0 & 0  \cr
 } } \right),
\end{eqnarray}
where $\gamma_{ij} = \Gamma_{\bot}^{ij}/\Gamma_{\|}$ is the phenomenological dephasing rate constants in units of the population relaxation rate constant.

At first, we consider the most simple case, when the line broadening is associated with the natural width only (the relaxation constants ratios yield $\gamma_{12}=\gamma_{13}=\gamma_{23}/2 = 1/2$). Substituting (\ref{eq:10}) into (\ref{eq:01}) gives the same complex set of differential equation, as in \cite{springerlink:10.1140/epjd/e2003-00080-2}, where use is made of the phenomenological approach  (${\bf{E}} \rightarrow {{\bf{E}}_{loc}}$) but with allowance for the quantum interference. For the numerical simulations we transform the obtained complex equations to the eight real ones (with the way analogues to Eqs.(\ref{eq:03})) to perform the numerical stability analysis of a steady-state solution and direct numerical integration of obtained real nonlinear equations.

\begin{figure}
\includegraphics[width=0.45\textwidth]{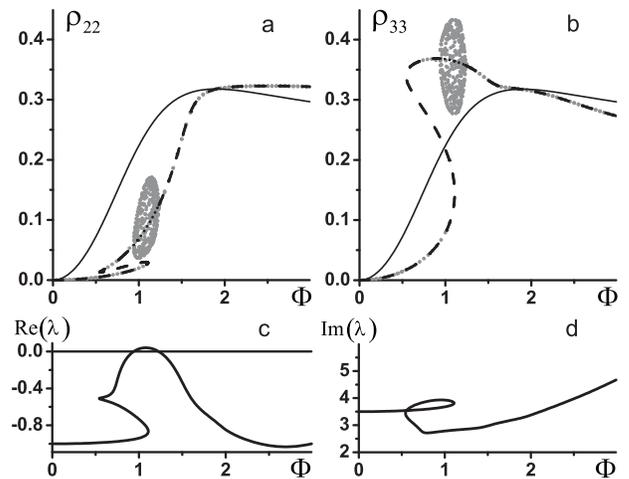}
\caption{Dynamical instability in (\ref{eq:01}) with operators (\ref{eq:10}) at the parameters $\delta=1.75$, $\beta=6$ and allowing the natural linewidth only. (a) and (b) show the dynamics of populations ${\rho _{22}}$ and ${\rho _{33}}$ where: the numerical solutions of the original differential equations are marked off by gray dots, the stable and unstable steady-state solutions are marked off by dashed and doted lines respectively, a thin solid line indicates the appropriate evolution of the energy level population but without NDD interaction. (c) shows the change in the sign of the real part of the complex eigenvalue (the associated imaginary part is shown in (d)).}\label{fig:01}
\end{figure}

Fig.~\ref{fig:01} shows stationary solutions of the energy level population ${\rho _{22}}$ and ${\rho _{33}}$ versus the Rabi frequency $\Phi$ at $\delta = \Delta/\Gamma_{\|}=1.75$, $\beta=\omega_{L}/\Gamma_{\|}=6$ and appropriate solutions of original differential equations after the integration time $\tau=100$. For comparison, the evolution of level populations at a negligibly small NDD interaction ($\beta=0$) is shown by a thin solid line. Fig.~\ref{fig:01} also demonstrates the evolution of the real and imaginary part of the complex eigenvalue obtained from the linear stability analysis of the steady-state solution. It is remarkable that there is an unstable region where there exists a pair of complex conjugated numbers with a positive real part. This is indicative of the Poincar\'e-Andronov-Hopf bifurcation occurrence, which predicts oscillations of the density matrix elements ${\rho _{ij}}$. Thus, we have answered the above-posed  question. That is, the  NDD interaction can lead to the dynamical instability and to this end, V-system-based media are quite promising. The calculation of the Lyapunov exponent spectrum inside the unstable region (it turns out to be $ \{{0,-,-,-,-,-,-,-}\} $) shows the periodicity of oscillations. In our consideration, the quantum interference is not taken into account $\left( {{\mbox{\boldmath$\mu$}_{21}} \cdot {\mbox{\boldmath$\mu$}_{31}}} \right) = 0 $ and in this case the effect predicted is most pronounced; otherwise oscillations are completely suppressed.

\begin{figure}
\includegraphics[width=0.45\textwidth]{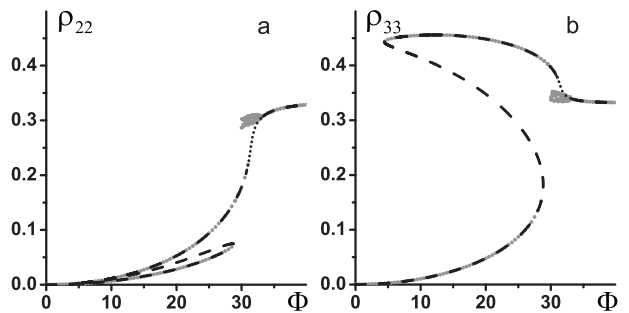}
\caption{The same as in Fig.~\ref{fig:01} (a) and (b) but for $\beta=500$, $\delta=55$, $\gamma_{12}=\gamma_{13}=10$ and $\gamma_{23}=4$. The parameters chosen are close to the oscillations threshold.} \label{fig:02}
\end{figure}

In the real systems the linewidth can exceed the natural linewidth due to the interaction of atoms with the surrounding environment and result in homogeneous and inhomogeneous broadening. The influence of homogeneous line broadening mechanism is shown on Fig.~\ref{fig:02}. As seen from the figure, the amplitude of osculations as well as their region are small. The increase in $\beta$ leads to the enhancement in amplitude and to the extension of the region of oscillations. According to calculations, the existence conditions of the predicted effect (similar to the IOB threshold) requires that $\omega_{L} > \sum_{ij} \Gamma_{\bot}^{ij}$ for the absence of the pure dephasing. With increasing the dephasing rate constant, the threshold condition of oscillations becomes $\omega_{L} \gg \sum_{ij} \Gamma_{\bot}^{ij}$ (compare the parameters in Fig.~\ref{fig:01} and Fig.~\ref{fig:02}). Estimations show that the threshold conditions can not be satisfied for $\gamma_{ij}>10^{2}$. So we can conclude that the existence condition of population oscillations is the comparability (within one order) of the relaxation rate constants. Further calculations show that any additional relaxation process in a system (relaxation between exited levels, collective relaxation effects) will merely suppress oscillations as an additional "friction" in a nonlinear system (this also follow from the basic principles of dynamics of nonlinear systems). The influence of the inhomogeneous broadening makes the problem much more complex although some preliminary conclusions can be made. It is expected that the inhomogeneous broadening can give rise to a more complex dynamics of the system and produce the chaotic behavior instead of the periodical one as well as suppress the effect, depending on the type of an inhomogeneous distribution function.

It remains for us to briefly discuss the feasibility of experimental observation of the predicted exhibition of the local-field effect. The predicted effect can take place in crystals doped by rare earth ions at concentration higher than $1~at.~\%$. In addition to the increase in the concentration, the homogeneous linewidth should be reduced by an appropriate choice of host media and cooling them to temperatures below $10 K$. This results in the comparable relaxation rates of population and polarization ($\Gamma_{\bot}/\Gamma_{\|} \approx 3 $ for $Pr^{3+}{:}Y_{2}SiO_{5}$) \cite{PhysRevB.52.3963}. Spectrally isolated three-level systems can be prepared with the help of spectral hole burning technics \cite{0953-4075-41-7-074001}. In so doing the inhomogeneous broadening is preferable to be minimized to increase the amount of ions in spectral holes.

The possibility of creating quantum dots (QD) arrays with low inhomogeneity ($< 1 meV$) \cite{SMLL:SMLL201000341} and the surface density up to $~10^{11}~cm^{-2}$ \cite{cheng:173108} allows one to propose an alternative system to observe oscillations. The V-type three-level system can be realized by fine-structure states of a single QD with orthogonal transitions at low temperatures ($ < 10~K$) where the linewidth is primarily lifetime-limited with $\Gamma_{\bot} \sim 10^{10}~Hz$ \cite{PhysRevLett.81.2759}. Single QD parameters of \cite{Bonadeo20111998,Gammon05071996,PhysRevLett.87.133603} were used to show the effect of population oscillations represented on Fig.~\ref{fig:01}. The estimations show that the effect is observable at the concentration higher than $10^{15}~cm^{-3}$. The tendency of creating QD arrays with higher homogeneity ($< 100 \mu eV$) which is required for the quantum computing \cite{SMLL:SMLL201000341} gives hope to reach the broadening ratio of order of units.

In conclusion, we have demonstrated the possibility of dynamical instabilities in dense media of resonant atoms with the number of energy levels greater than two (this is exemplified by the V-type three-level atom model) and suggested some credible systems where their manifestations can occur. 

\begin{acknowledgments}
The work was supported by Grants RFFI/BRFFI (F10R-170 and 10-02-90047-Bel\_a).
\end{acknowledgments}

\bibliography{article}

\end{document}